\documentclass[10pt,conference,letterpaper]{IEEEtran}
\makeatletter
\def\ps@headings{%
\def\@oddhead{\mbox{}\scriptsize\rightmark \hfil \thepage}%
\def\@evenhead{\scriptsize\thepage \hfil \leftmark\mbox{}}%
\def\@oddfoot{}%
\def\@evenfoot{}}
\makeatother
\usepackage{graphicx,cite}

\usepackage{amssymb,amsmath,amsbsy}

\usepackage{algorithm,algorithmic}

\usepackage{array}

\usepackage{subfigure}

\usepackage{balance}

\usepackage[hyphens]{url}
\usepackage{hyperref}

\usepackage{multirow}

\def\squareforqed{\hbox{\rlap{$\sqcap$}$\sqcup$}}
\def\qed{\ifmmode\squareforqed\else{\unskip\nobreak\hfil
\penalty50\hskip1em\null\nobreak\hfil\squareforqed
\parfillskip=0pt\finalhyphendemerits=0\endgraf}\fi}

\begin{document}

\title{Low Latency Datacenter Networking: \\A Short Survey}
\author{\IEEEauthorblockN{Shuhao Liu\IEEEauthorrefmark{1}\IEEEauthorrefmark{2}, Hong Xu\IEEEauthorrefmark{1}, Zhiping Cai\IEEEauthorrefmark{2}}
\IEEEauthorblockA{ 
\IEEEauthorrefmark{1}Department of Computer Science, City University of Hong Kong
}
\IEEEauthorblockA{ 
\IEEEauthorrefmark{2}College of Computer, National University of Defence Technology
}
}
\maketitle


\begin{abstract}

Datacenters are the cornerstone of the big data infrastructure supporting numerous online services. The demand for interactivity, which significantly impacts user experience and provider revenue, is translated into stringent timing requirements for flows in datacenter networks. Thus low latency networking is becoming a major concern of both industry and academia.

We provide a short survey of recent progress made by the networking community for low latency datacenter networks. We propose a taxonomy to categorize existing work based on four main techniques, reducing queue length, accelerating retransmissions, prioritizing mice flows, and exploiting multi-path. Then we review select papers, highlight the principal ideas, and discuss their pros and cons. We also present our perspectives of the research challenges and opportunities, hoping to aspire more future work in this space.

\end{abstract}

\section{Introduction}
\label{sec:intro}

Datacenters with hundreds of thousands of servers are the powerhouse for numerous Internet applications and services today. Many of them, such as search, online shopping, and social networking, are interactive in nature with stringent latency requirements. The response to a user's request must be delivered fast enough, and in fact even a 100~ms latency overhead greatly degrades user experience and provider revenue \cite{Zats-2012-DeTail}. 

In this paper, we focus on the latency issue {\em inside} a datacenter network, the critical underpinning of the big data infrastructure that moves bits and bytes across the boundary of servers. As Internet services popularize, providers build large-scale distributed systems to store and process the big data. Thus, many applications are designed following a partition-aggregation pattern \cite{Alizadeh-2010-Data,Zats-2012-DeTail}: The processing of a request is divided into many small tasks, each of which handled by an individual server, and the final response is assembled from the results of these tasks. The results are transmitted as TCP flows that are usually small in size \cite{Alizadeh-2010-Data,Alizadeh-2013-pFabric,Zats-2012-DeTail}. As a result, network latency---the time it takes for the network to transmit all the results for aggregation---constitutes a major part of the user-perceived response latency.

It is well-known that current datacenter network design does not provide adequate latency performance, especially for short TCP flows. The average flow completion time (FCT) is 2x--3x its theoretical minimum \cite{Alizadeh-2010-Data,Alizadeh-2013-pFabric}. Moreover, the tail FCT, e.g. the 99-th or 99.9-th percentile FCT, can be more than 10x larger than the average \cite{Zats-2012-DeTail,Alizadeh-2013-pFabric}. Tail latency is crucial for applications with partition-aggregation workflows, since the response needs to wait for the slowest flow to finish.

The major culprit of high latency is long queueing delay in switches, caused by the traffic characteristics of datacenters. Specifically, datacenter networks carry two types of flows: {\em mice flows} that are latency-sensitive short requests and responses to support interactive applications, and {\em elephant flows} that transfer bulky traffic for data replication and data shuffling in the intermediate stage of parallel computation jobs \cite{Alizadeh-2010-Data,Zats-2012-DeTail}. Though few in number, elephant flows account for the majority of the bytes \cite{Alizadeh-2010-Data}. Thus when a mice flow enters a switch, very likely it will be queued behind a number of packets from elephant flows. Since mice flows have only a few packets, queueing delay becomes a significant portion of the total transmission delay.

Our community has recognized low latency networking as an important research problem, and devoted much attention to tackle the issue from various perspectives. The objective of this survey is to provide a concise account of the initial progress towards this quest, which to our knowledge has not been done before. It is our view that in general, state-of-the-art research focuses on four main techniques: reducing queue length, accelerating retransmissions, prioritizing mice flows, and exploiting multipath. Targeting at reducing queueing delay and transmission timeouts, existing efforts either attack the problem head-on by reducing the queue occupancy and accelerating retransmissions, or indirectly by changing the packet scheduling discipline to prioritize mice flows, and by intelligently choosing the right path to circumvent congestion.

This survey is organized as follows. We first present an overview of existing research in Sec.~\ref{sec:taxonomy}. Then, we examine individual work in detail in Sec.~\ref{sec:reducequeue}--\ref{sec:multipath} according to the four-category taxonomy. We outline challenges and opportunities in Sec.~\ref{sec:discussion}, and conclude this paper in Sec.~\ref{sec:conclusion}.

We emphasize that because of the space constraint, we only sample a small number of papers here \cite{Alizadeh-2010-Data,Alizadeh-2012-Less,Wilson-2011-Better,Hong-2012-Finishing,Alizadeh-2013-pFabric,Zats-2012-DeTail,Xu-2013-RepFlow}. The survey is by no means comprehensive in the sense of reflecting the landscape of low latency networking, not to mention related research efforts in systems and other communities. Thus, this survey only reflects the authors' perspective and ignorance.

\section{A Taxonomy}
\label{sec:taxonomy}


\begin{table*}
\centering
\caption{An Overview of Low Latency Datacenter Networking Proposals}
\centering
\begin{tabular}{c|c|c|c|c|c|c|c}
  \hline
  \multicolumn{2}{c|}{\multirow{2}{*}{Categories}} & \multirow{2}{*}{Proposals} & \multicolumn{2}{c|}{Objectives}
        & \multicolumn{3}{c}{Modifications to} \\
        \cline{4-8} \multicolumn{2}{c|}{} & &  FCT & deadline & TCP & switches & applications \\
  \hline
  \multicolumn{2}{c|}{\multirow{2}{*}{Reducing queue length}}
        & DCTCP \cite{Alizadeh-2010-Data} & mean & $\times$ & $\surd$ & $\times$ & $\times$ \\
        \cline{3-8}
        \multicolumn{2}{c|}{} & HULL \cite{Alizadeh-2012-Less} & mean & $\times$ & $\surd$ & $\surd$ & $\times$ \\
  \hline
  \multirow{4}{*}{} & \multirow{4}{*}{Flow deadlines}
        & D$^3$ \cite{Wilson-2011-Better} & none & $\surd$ & $\surd$ & $\surd$ & $\times$ \\
        \cline{3-8}
        & & PDQ \cite{Hong-2012-Finishing} & mean & $\surd$ & $\surd$ & $\surd$ & $\times$ \\
        \cline{3-8}
         Prioritizing flows & & D$^2$TCP \cite{Vamanan-2012-Deadline} & tail & $\surd$ & $\surd$ & $\times$ & $\times$ \\
         \cline{3-8}
         based on & & MCP \cite{Chen-2013-Towards} & tail & $\surd$ & $\surd$ & $\times$ & $\times$ \\
        \cline{2-8}
     & \multirow{2}{*}{Application assignment}
        & pFabric \cite{Alizadeh-2013-pFabric} & mean \& tail & $\surd$ & $\surd$ & $\surd$ & $\surd$ \\
        \cline{3-8}
        & & \multirow{2}{*}{DeTail \cite{Zats-2012-DeTail}} & \multirow{2}{*}{tail} & \multirow{2}{*}{$\surd$} & \multirow{2}{*}{$\surd$} & \multirow{2}{*}{$\surd$} &  \multirow{2}{*}{$\surd$} \\
        \cline{1-2}
  \multicolumn{2}{c|}{\multirow{2}{*}{Exploiting multipath}}
        & & & & &  & \\
        \cline{3-8}
         \multicolumn{2}{c|}{} & RepFlow \cite{Xu-2013-RepFlow} & mean \& tail & $\times$ & $\times$ & $\times$ & $\surd$ \\
  \hline
    \multicolumn{2}{c|}{\multirow{3}{*}{Accelerating retransmissions}}
        & DIBS \cite{Zarifis-2014-DIBS} & tail & $\times$ & $\surd$ & $\surd$ & $\times$ \\
        \cline{3-8}
    \multicolumn{2}{c|}{} & FastLane \cite{Zats-2013-FastLane} & tail & $\times$ & $\surd$ & $\surd$ & $\times$ \\
        \cline{3-8}
    \multicolumn{2}{c|}{} & CP \cite{Cheng-2014-CP} & tail & $\times$ & $\surd$ & $\surd$ & $\times$ \\
    \hline
\end{tabular}
\label{tab:overview}
\end{table*}
Queueing delay is the culprit of high latency in datacenter networks. According to measurement results \cite{Alizadeh-2010-Data}, even with mild congestion, queueing delay is a significant part of FCT for mice flows. To reduce queueing delay and improve latency, researchers have mainly exploit four techniques: reducing queue length, accelerating retransmissions, prioritizing mice flows, and exploiting multipath.

Reducing switch queue length, or buffer occupancy, is the most direct way to tackle the latency problem. Commodity switches use deep buffers to handle bursty traffic. In datacenter networks where the delay-bandwidth product is small, deep buffers can be detrimental to latency performance. As TCP relies on packet drops for congestion control, elephants will keep sending before the buffer is filled up, yielding long queueing delay for mice. Hence, one line of work targets at reducing the queue length in switch buffers, with novel rate control and congestion detection mechanisms at the transport layer that require modifications to both end-hosts and switches. We discuss two representative works, DCTCP \cite{Alizadeh-2010-Data} and HULL \cite{Alizadeh-2012-Less}, in Sec.~\ref{sec:reducequeue}.

Another intuitive idea is to reduce delay caused by packet losses, as TCP performs retransmissions after an RTO. Provided that an RTO is usually much larger than the end-to-end RTT, waiting for an RTO causes unnecessary delay. Some have proposed to reduce or avoid retransmission delay either by generating explicit notifications about dropped packets \cite{Zats-2013-FastLane, Cheng-2014-CP}, or by exploiting the idle paths to help absorb them \cite{Zarifis-2014-DIBS}.

The third line of work employs priority scheduling (Sec.~\ref{sec:priority}). Instead of treating each packet equally, switches prioritize packets of certain mice flows and transmit them before others in the queue, thereby greatly improve their FCT. There are two ways to prioritize flows. First, we can ask applications to explicitly assign priority \cite{Alizadeh-2013-pFabric,Zats-2012-DeTail}. Alternatively, we can utilize the deadline information available for flows of certain applications today as the implicit priority. Some applications associate a deadline of around 200~ms--300~ms to their flows, and those that did not make it are simply discarded to guarantee the response time \cite{Wilson-2011-Better,Hong-2012-Finishing,Vamanan-2012-Deadline,Chen-2013-Towards}. Therefore, prioritizing flows through in-network scheduling is an effective method, possibly with help from a revamped transport layer and link layer.

The fourth approach exploits the inherent multi-path nature of datacenter networks. Many network topologies provide rich connectivities across servers, with many paths of equal hop count. Current multi-path solution is ECMP (Equal Cost Multi-path) that uniformly at random distributes flows over the network based on the hash of a flow's five tuple. This is sub-optimal as congestion is typically localized. By choosing a less congested path, we can reduce queueing delay experienced by the flow. Therefore, researchers have proposed various multi-path load balancing schemes ranging from simple flow replication \cite{Xu-2013-RepFlow} to per-packet adaptive path selection \cite{Zats-2012-DeTail}.

Note that these techniques are not mutually exclusive. In fact, some schemes integrate multiple techniques for greater performance gains. For example, DeTail \cite{Zats-2012-DeTail} leverages both priority scheduling and congestion-aware multi-pathing. In addition to the four-cateogry taxonomy, existing work also differs in many other aspects. Table~\ref{tab:overview} presents a more detailed comparison.

\section{Reducing Queue Length}
\label{sec:reducequeue}

DCTCP \cite{Alizadeh-2010-Data} and HULL \cite{Alizadeh-2012-Less} are two proposals that strive to keep the switch buffer occupancy persistently low to reduce latency.

\subsection{DCTCP}
\label{sec:dctcp}

DCTCP (Data Center TCP) \cite{Alizadeh-2010-Data} is arguably the first work targeting at reducing latency in datacenter networks. It lays down the groundwork by making two seminal contributions. Alizadeh et al. \cite{Alizadeh-2010-Data} first study the typical application needs and traffic characteristics in production datacenters through measurements. The traces collected over a one-month period show that, most flows are delay-sensitive short flows, which experience long latency due to elephant flows occupying some or all of the buffer in switches \cite{Alizadeh-2010-Data}. Motivated by these observations, the authors design a new transport protocol, DCTCP, that reacts to congestion {\em in proportion to} the extend of congestion in order to reduce buffer occupancy without much loss of throughput.

\begin{figure}
  \centering
  \includegraphics[width=.4\textwidth]{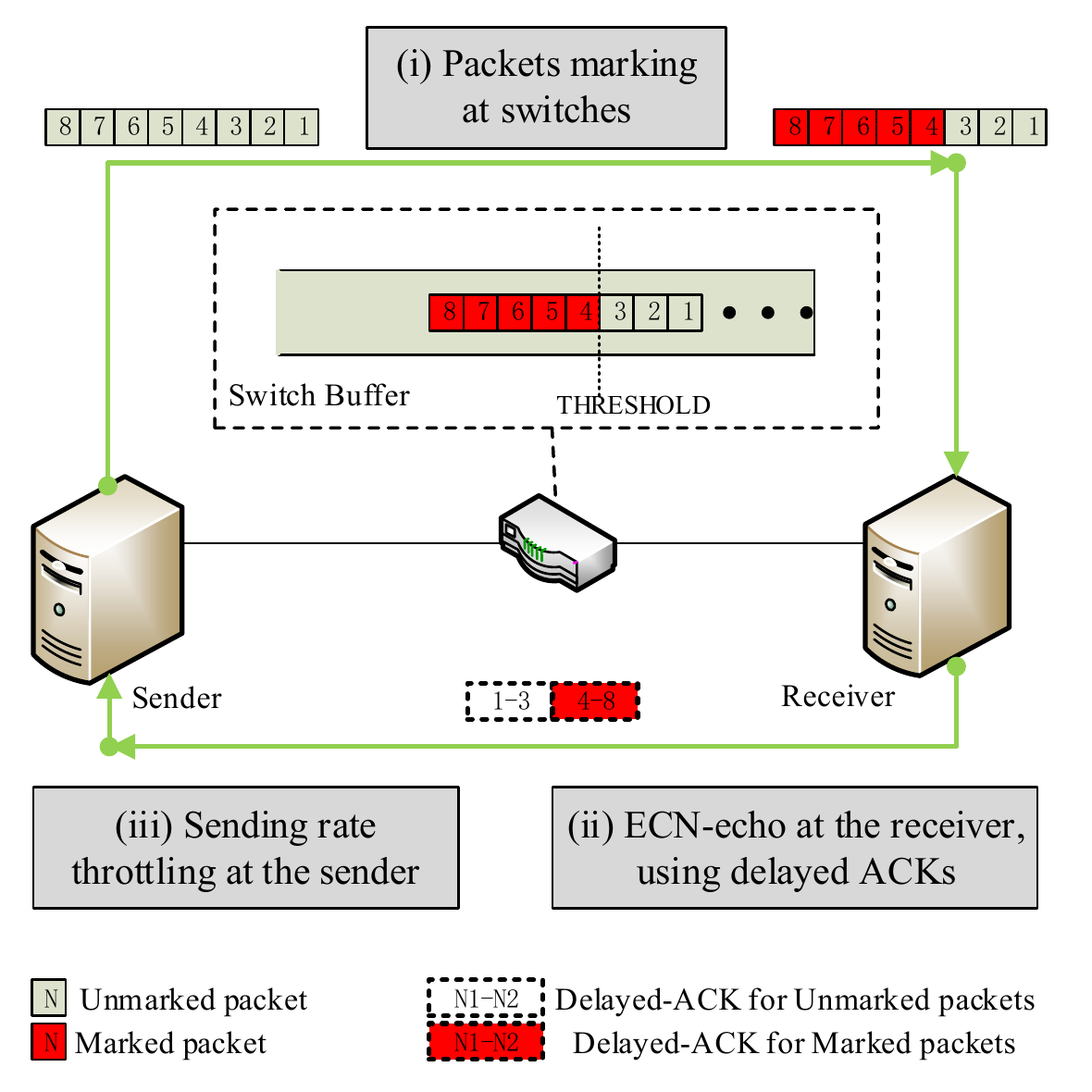}\\
  \caption{The control loop in DCTCP.} 
  \label{fig:dctcp}
  \vspace{-3mm}
\end{figure}

The key design of DCTCP is summarized in Fig.~\ref{fig:dctcp}. DCTCP relies on Explicit Congestion Notification (ECN), a feature commonly available on commodity switches. Whenever the buffer occupancy exceeds a small threshold, switches start to mark a one-bit field in the incoming packets. This information is relayed back to the sender in the delayed ACKs with the ECN-echo flag set by the DCTCP receiver. 
More importantly, the DCTCP sender maintains an estimate of the fraction of ECN marked packets, and reduces the window size {\em proportionally}: the larger the fraction, the more the congestion window would back off. As a result,
the sender is able to detect congestion {\em before} the queue starts to build up, thus ensuring low buffer occupancy.

DCTCP's new control logic requires only 30 lines of code change to legacy TCP, making it practical for deployment in current infrastructures. A testbed implementation shows that DCTCP improves the FCT greatly, especially at the tail: the 99.9-th percentile FCT is reduced by 40\% over TCP.

\subsection{HULL}

HULL (High-bandwidth Ultra-Low Latency) \cite{Alizadeh-2012-Less} is an incremental work based on DCTCP. It focuses on \emph{eliminating} queues at all ports to further reduce the queue length. In DCTCP, small queues are still needed to detect congestion. HULL is motivated by the idea that it is possible to eliminate buffering and queueing delay by detecting congestion based on the link utilization approaching its capacity, rather than the queue occupancy. 

Specifically, HULL utilizes {\em phantom queue}, a virtual queue maintained at each port that sets ECN marks based on link utilization. It simulates queue buildup for a virtual link running at a configurable speed {\em slower} than the physical capacity, without buffering anything. When the phantom queue is above a threshold, the corresponding port marks incoming packets with ECN which is utilized by DCTCP to perform adaptive congestion control. Since the phantom queue caps the aggregated flow rate to less than the physical capacity, the buffers are kept largely unoccupied, and packets experience baseline transmission delay without queueing.
The authors show that HULL can reduce both average and 99-th percentile latency by more than 10x over DCTCP and 40x over TCP, with a 10\% bandwidth reduction as the tradeoff.


\section{Accelerating Retransmissions}

When mice flows coexist with elephants, packet drops and retransmissions are inevitable. The retransmission delay to recover packet drops thus plays a critical role in determining the tail latency. In TCP, the FCT of a mice with packet drops will not be less than one RTO, since the sender waits for one RTO before retransmitting. However, RTO needs to be larger than the largest possible RTT to confirm packet drops, which is too long for a flow to meet its deadline. Another consequence of packet drops is they reduce TCP congestion window size, and more round trips are needed to complete the flow. 

Existing work solves this problem from two perspectives. DIBS \cite{Zarifis-2014-DIBS} opportunistically re-directs a packet from a congested switch port to another at random to avoid packet drops and retransmission. FastLane \cite{Zats-2013-FastLane} and CP (Cutting Payload) \cite{Cheng-2014-CP}, on the other hand, aim to accelerate retransmissions by generating fast and explicit notifications.

\subsection{DIBS}


Detour-Induced Buffer Sharing \cite{Zarifis-2014-DIBS}, or DIBS in short, makes use of idle network resources to absorb traffic bursts at hot spots. When a packet has to be dropped, the switch forwards it to another random port instead, in the hope that it can eventually be delivered. There are three possible outcomes: i) The packet is delivered via another path with more hops, making a detour around the hot spot; ii) The packet loops back to the same egress port, finding its buffer no longer full, and is transmitted through; and iii) The packet does not find an idle path after reaching its TTL and is eventually dropped.
Owing to the multi-path nature of datacenter networks, a little detour is likely to avoid most of the packet drops. Only the third case, which is proved to be highly unlikely, will trigger timeout and retransmission. 

\subsection{FastLane and CP}

FastLane \cite{Zats-2013-FastLane} and CP \cite{Cheng-2014-CP} propose to generate explicit loss notifications in order to accelerate retransmissions---the delay to trigger retransmissions is reduced from one RTO to one RTT. 

FastLane \cite{Zats-2013-FastLane} generates a new notification packet at the switch where a packet is dropped, and sends it back to the sender immediately. To minimize CPU overhead, the switch generates the notification by simply swapping the source and destination addresses in the IP header of the dropped packet. Also, a bandwidth cap may be enforced to reduce the bandwidth overhead.

On the other hand, CP \cite{Cheng-2014-CP} uses the original header of the dropped packet as the notification. Only the bulky payload is removed when the packet is dropped. The receiver recognizes the missing payload upon receiving the header, and triggers retransmission using a SACK-like precise ACK.
Both FastLane and CP are compatible with any TCP protocol. 

\section{Prioritizing Mice Flows}
\label{sec:priority}

Both of the above approaches treat each packet equally, no matter whether it belongs to an elephant flow or a mice. While this is in general acceptable to ensure fairness in the Internet, some argue for an {\em unfair} treatment of flows in datacenters, where mice flows should be prioritized so they are not blocked by elephants \cite{Wilson-2011-Better,Hong-2012-Finishing,Alizadeh-2013-pFabric, Zats-2012-DeTail}. Here we discuss this line of work in more detail.
Particularly, we categorize the proposals into two kinds depending on the source of the priority information: one that uses the deadline associated with the flow to prioritize, e.g. D$^3$ \cite{Wilson-2011-Better} and PDQ \cite{Hong-2012-Finishing}, and another that assumes priority is explicitly assigned by applications, e.g. pFabric \cite{Alizadeh-2013-pFabric} and DeTail \cite{Zats-2012-DeTail}.


\subsection{Deadline as priority}

\subsubsection{D$^3$}
\label{sec:D3}

D$^3$ (Deadline Driven Delivery) \cite{Wilson-2011-Better} is an early attempt to perform deadline-aware flow prioritization. The intuition behind D$^3$ is that the network should try its best to meet a flow's deadline by calculating and allocating enough bandwidth for it, i.e., to arbitrate the flow sending rate \cite{Munir-2014-Friends}.


Specifically, Wilson et al. \cite{Wilson-2011-Better} argue that many interactive applications have deadlines for flows initiated by the worker machines of the application. Using this information, a flow in D$^3$ calculates the minimum rate required to transmit pending data in time, and embed it in the packet header. The flow thus ``requests'' bandwidth allocation across all the switches along the path. The switches respond by notifying the sender about the amount of reserved bandwidth via ACKs. The sender adjusts the flow rate and starts the next iteration of bandwidth allocation request.


A switch in D$^3$ performs centralized bandwidth allocation based on global knowledge of active flows. Limited by the compute and memory capacities, a switch allocates bandwidth in a first-come-first-serve (FCFS) manner. Thus, it may only satisfy the first few flows' demands when there is insufficient bandwidth for all. The effectiveness of D$^3$ is confirmed by an implementation on a small-scale testbed. 


\subsubsection{PDQ}
\label{sec:PDQ}

\begin{figure}
  \subfigure[Flow information]{
    \begin{minipage}[b]{0.5\textwidth}
      \centering
\begin{footnotesize}
      \begin{tabular}{c|ccc}
        & Size & Deadline & T$_{arrival}$ \\ \hline
        A & 1 & 3 & 1 \\
        B & 2 & 5 & 2 \\
        C & 5 & 9 & 0 \\
        \multicolumn{2}{c}{}\\
      \end{tabular}
\end{footnotesize}
    \end{minipage}}
  \subfigure[DCTCP \cite{Alizadeh-2010-Data}: A misses its deadline. Bandwidth is allocated equally among all requests.]{
    \centering
    \begin{minipage}[b]{0.48\textwidth}
      \centering
      \includegraphics[width=.7\textwidth, trim = 0 115 0 0, clip]{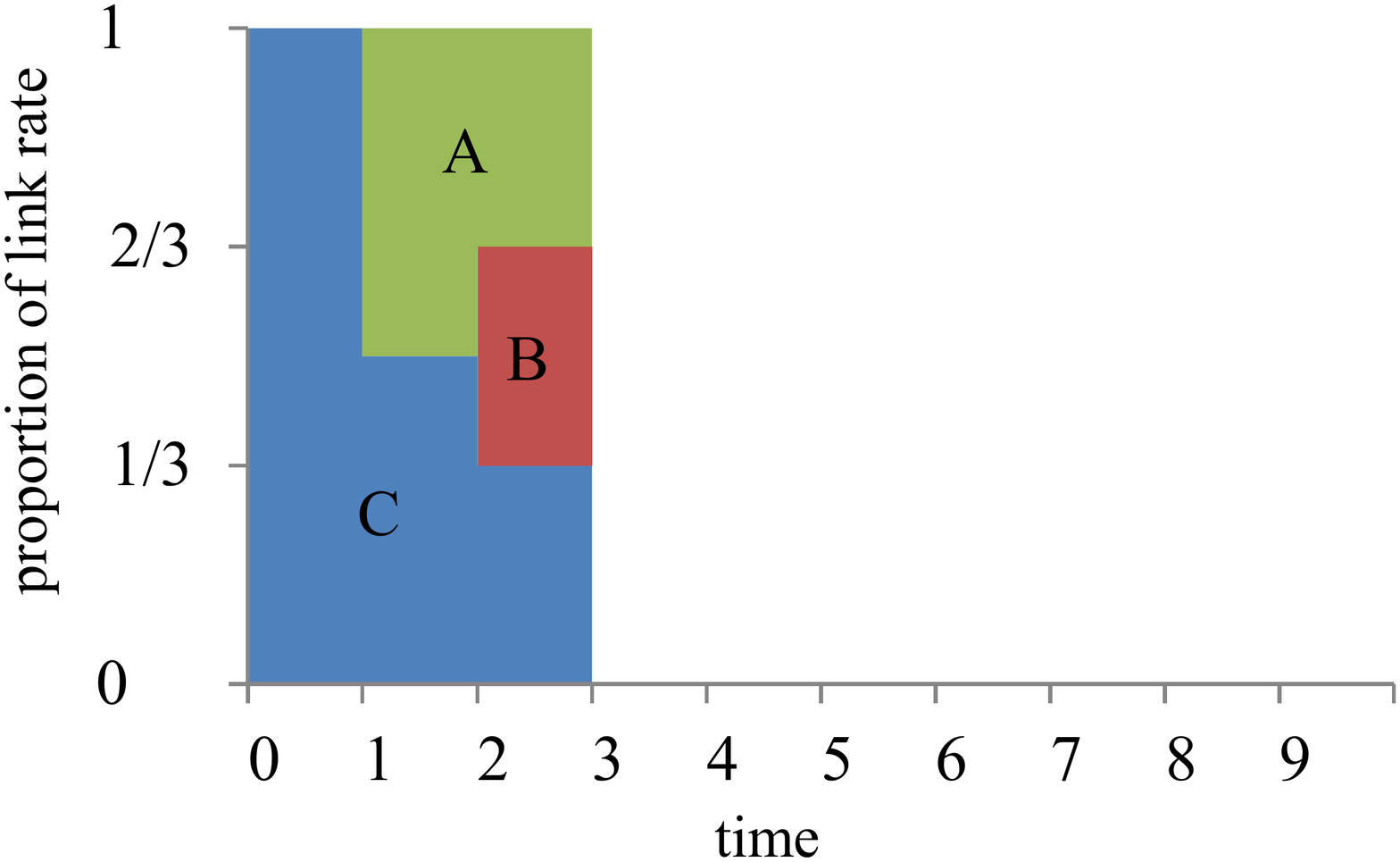}
    \end{minipage}}
  \subfigure[D$^3$ \cite{Wilson-2011-Better} (for service order A$>$C$>$B): B misses its deadline. In each allocation loop, C's request is served earlier than B, so that C always gets its desired rate (1/2 link rate). Then, B is allocated the rest of available bandwidth, which is less than its desired rate (2/3 link rate). However, if B could be served prior to C, its deadline would be satisfied.]{
    \centering
    \begin{minipage}[b]{0.48\textwidth}
      \centering
      \includegraphics[width=.7\textwidth, trim = 0 115 0 0, clip]{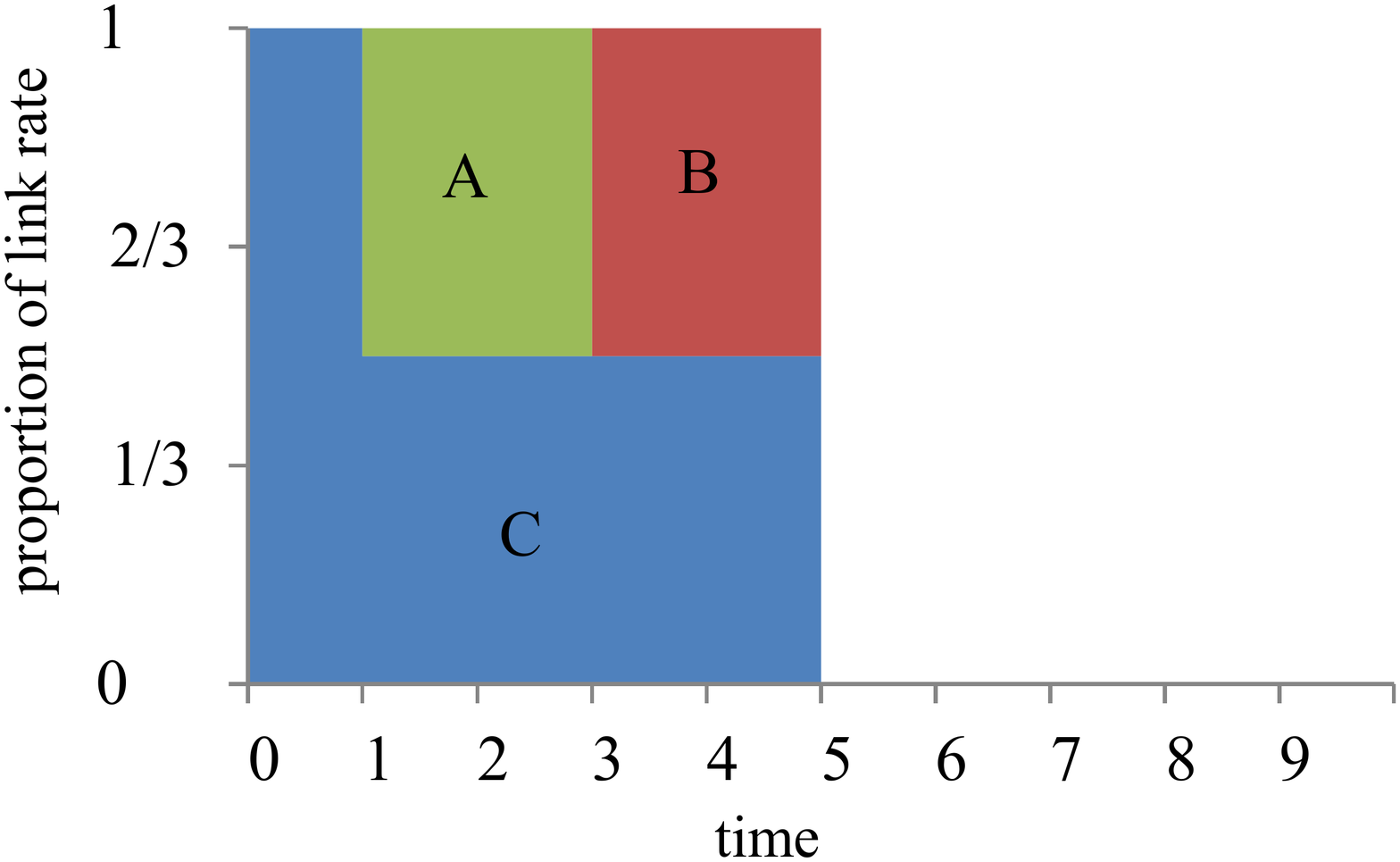}
    \end{minipage}}
      \subfigure[PDQ \cite{Hong-2012-Finishing} (ideal performance): When $time = 1$, A preempts C for having an earlier deadline. When $time = 2$, A is completed, and B preempts C for the same reason. A and B are both able to complete as quickly as possible.]{
    \label{fig:pdq} 
    \centering
    \begin{minipage}[b]{0.48\textwidth}
      \centering
      \includegraphics[width=.7\textwidth, trim = 0 115 0 0, clip]{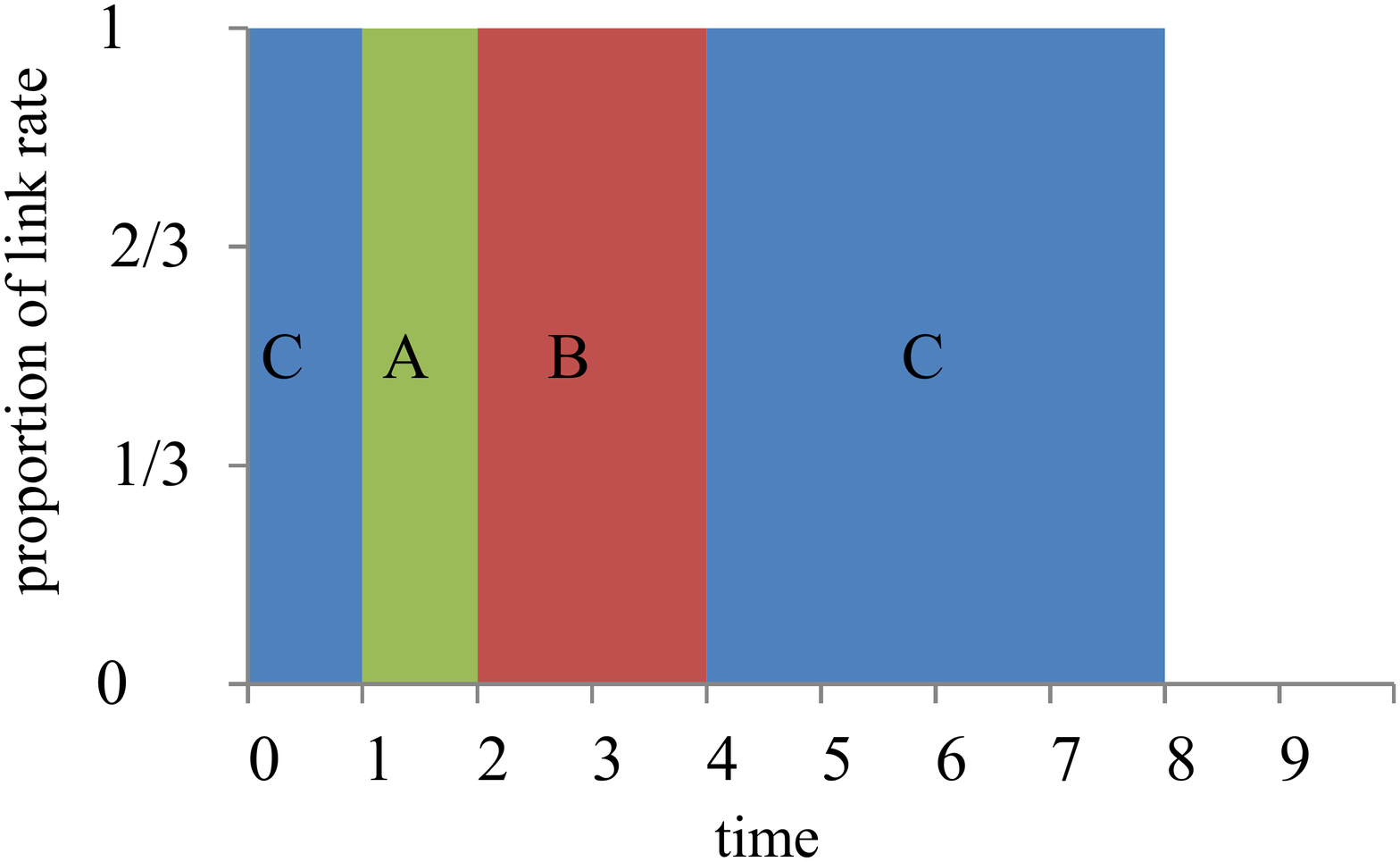}
    \end{minipage}}
  \caption{Fair-share (e.g. DCTCP) and D$^3$ are sub-optimal in meeting deadlines.}
  \label{fig:pdqmotivation} 
  \vspace{-3mm}
\end{figure}

D$^3$'s FCFS scheduler may incur performance issues. An example is illustrated in Fig.~\ref{fig:pdqmotivation}. This motivates the design of PDQ, Preemptive Distributed Quick flow scheduling \cite{Hong-2012-Finishing}. In a nutshell, PDQ improves upon D$^3$ by allowing {\em preemptive scheduling}, and by proactively giving bandwidth to the most critical flows as illustrated in Fig.~\ref{fig:pdq}. 

PDQ builds on two real-time job scheduling disciplines: Earliest Deadline First (EDF) and Shortest Job First (SJF). EDF and SJF are centralized and difficult to apply to scheduling flows. PDQ provides a distributed scheduling layer to allow switches to collaborate and converge to stable allocation decisions. This primitive is used to approximate EDF and SJF effectively using only FIFO droptail queues. Simulation study using real-world workloads shows that PDQ can reduce mean FCT by $\sim$30\% over D$^3$.




\subsubsection{D$^2$TCP and MCP}

Adjusting TCP window size is a classic way of congestion control. Motivated by DCTCP \cite{Alizadeh-2010-Data} which exploits ECN feedback to estimate the extent of congestion, D$^2$TCP \cite{Vamanan-2012-Deadline} and MCP \cite{Chen-2013-Towards} further take flow deadlines into account.

As an earlier proposal, D$^2$TCP \cite{Vamanan-2012-Deadline} uses a gamma-correction function to adaptively modulate TCP window size based on flow deadlines. That is, near-deadline flows get a larger window than others even if their congestion estimations are the same. MCP \cite{Chen-2013-Towards} takes one step further, arguing that the window size should be modulated to meet the deadline from the very beginning of transmission. It sets a theoretical foundation for ECN-based congestion control, formulating it as an optimization problem which aims at minimizing the long-term average per-packet delay. The deadline guarantees and network stability are modeled as constraints. Chen et al. solve for the optimal window size by converting it to a convex program, and propose an approximation algorithm which is shown to be effective through numerical evaluation.

\subsection{Priority assigned by applications}

We now discuss proposals for which a flow's priority is explicitly assigned by its application.

\subsubsection{DeTail}
\label{sec:detail2}

DeTail \cite{Zats-2012-DeTail} is a cross-layer framework designed to reduce the tail FCT. Zats et al. identify two main causes of long tail latency, and develop two techniques to mitigates them. The first cause is the absence of prioritization among flows. During flash congestion, mice flows are stuck behind elephant flows, causing them to hit the long tail. DeTail utilizes the recently standardized Priority Flow Control (PFC) at the link layer to attack this problem. PFC is available on newer Ethernet switches, and employs a priority based pausing mechanism that where switches notify the senders to temporarily halts the transmission of low-priority flows during congestion.


The second cause of long tail latency is uneven load balancing at the network layer. Current hash-based load balancing schemes may direct several elephant flows to the same path despite the availability of less congested paths. DeTail adopts an adaptive multipath load balancing scheme to mitigate this problem, whose discussion is postponed to Sec.~\ref{sec:detail}. 


\subsubsection{pFabric}
\label{sec:pfabric}

Unlike most work that integrates several mechanisms to build a complex system, pFabric \cite{Alizadeh-2013-pFabric} follows a minimalism design in tackling the latency issue.

The principle of pFabric is to maximize the benefits of flow prioritization. Alizadeh et al. assert that differentiating flow criticality via rate control, though widely used, is neither effective nor easy to implement. Instead of throttling or dampening flow rates at senders with local and greedy decisions, it is good enough to have the switches {\em alone} prioritize the mice flows, while all the flows are simply transmitting at line rate without (complicated) rate control. 

pFabric is a clean-slate network stack designed with the above insight. Mice flows that are near-deadline and more critical to user experience carry a small priority value in their packet headers. Packets are then sorted in the non-descending order of their priority values in every switch queue, so that higher-priority flows always get transmitted first. Moreover, switches are extremely shallow-buffered. A buffer is easily filled up and starts to drop packets in the non-ascending order of priority. Packet-level ns-2 simulations demonstrate that pFabric achieves near-optimal latency performance both in the average and in the tail.



\section{Utilizing Multi-path}
\label{sec:multipath}

As mentioned earlier, hash based multipath routing such as ECMP causes hash collision among flows and results in long queueing delay for mice flows. In this section, we present two proposals that focus on choosing the right path to circumvent congestion, in addition to reducing congestion on each path directly.


\subsection{DeTail}
\label{sec:detail}

\begin{figure}
  \centering
  \includegraphics[width=.5\textwidth]{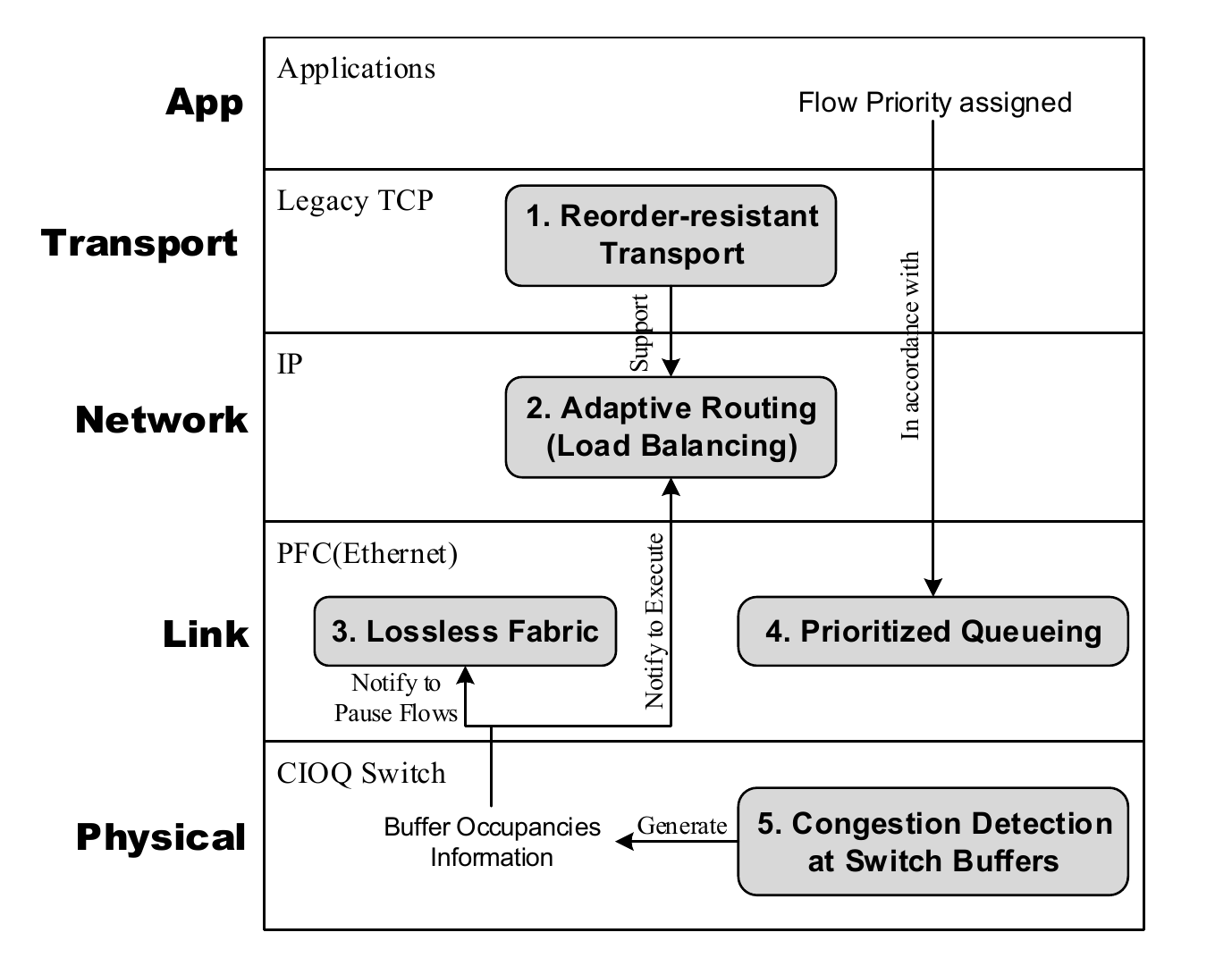}\\
  \caption{DeTail's cross-layer design, with changes to the protocol stack and the system information flow. } \label{fig:detail}
\end{figure}

As mentioned in Sec.~\ref{sec:detail2}, DeTail \cite{Zats-2012-DeTail} proposes an adaptive load balancing mechanism at the network layer to better utilize multiple paths in the network. Together with flow prioritization at the link layer, these two constitute the complete cross-layer design as depicted in Fig.~\ref{fig:detail}.

Essentially, DeTail's adaptive load balancing makes per-packet congestion-based routing decisions: when a packets arrives at a switch, it is forwarded to one of the shortest paths that are underutilized. 
If the current path is congested, the PFC pause message sent by the switch immediately triggers a route change. In case no idle path is available, the flow is simply paused. Since the mechanism works on a per-packet basis, DeTail disables the monitoring and reaction to out-of-order packet delivery in legacy TCP as shown in Fig.~\ref{fig:detail}. 





\subsection{RepFlow}

Previous work on reducing latency requires modifications to switches, end-host network stack, or even the network fabric itself. RepFlow \cite{Xu-2013-RepFlow} is a simple yet effective approach that replicates each mice flow to reduce FCT, without any change to switches or end-host kernels.

\begin{figure}
  \centering
  \includegraphics[width=.35\textwidth]{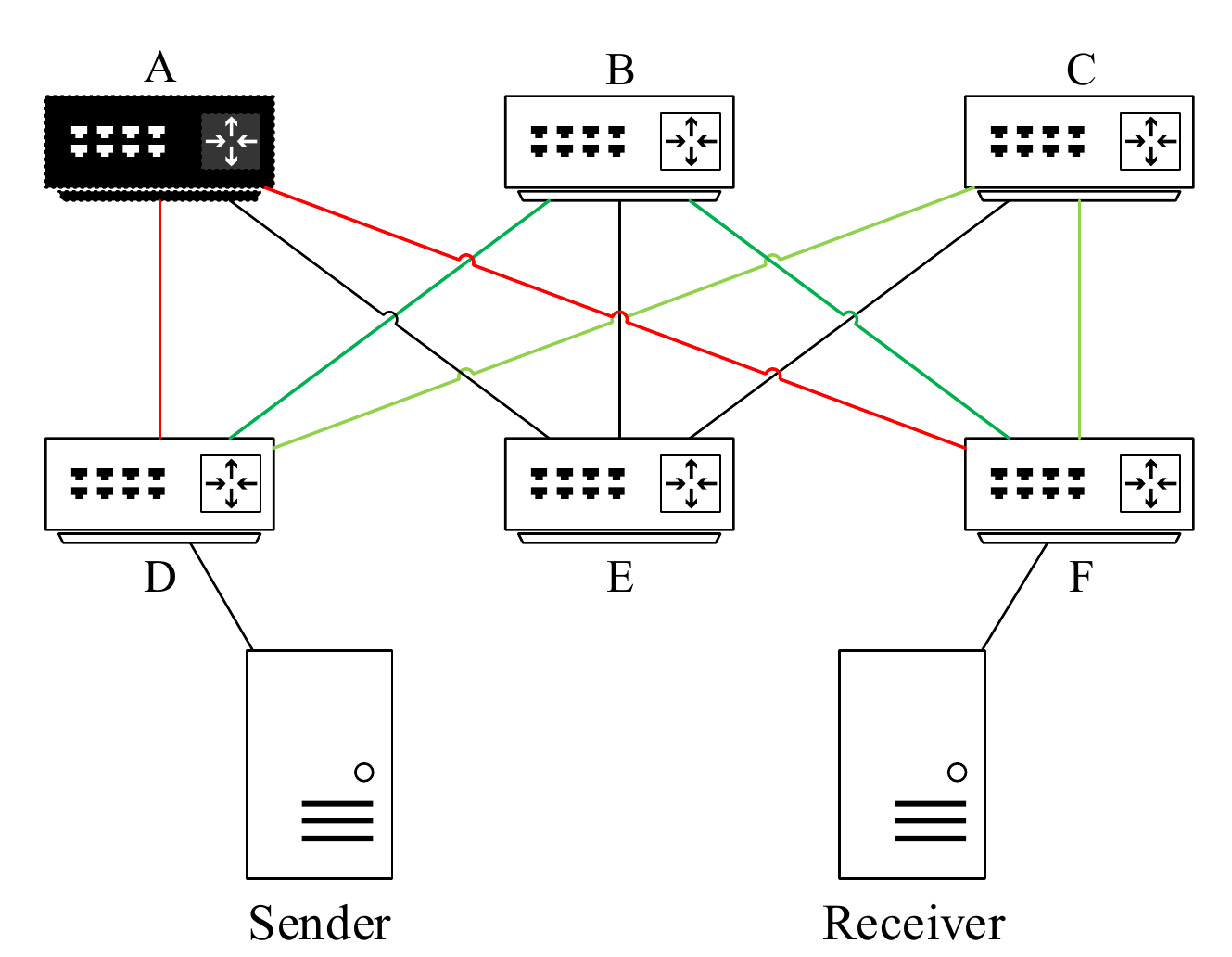}\\
  \caption{An example to understand RepFlow. With ECMP, one of the three colored paths are chosen with equal probability. Assuming that the black switch is suffering from congestion caused by other elephant flows, the flow has a probability of 1/3 to experience long queueing delay. With replication, at least one of the copies traverses a path without the congested switch.} \label{fig:repflow}
\end{figure}

The key insight behind RepFlow is that multi-path diversity, which is readily available with datacenter network topologies such as fat-tree, is an effective means to combat long queueing delay. Flash congestion due to bursty traffic and ECMP hash collision happen randomly in any part of the network at any time. As a result, congestion levels on different paths are statistically independent. In RepFlow, the replicated and the original flows are highly likely to traverse different paths, and the probability that {\em both} experience long queueing delay is much smaller. Fig.~\ref{fig:repflow} shows an example of the intuition.


The traffic characteristics of datacenter networks lend well to RepFlow. Though more than 90\% of all the flows are mice flows ($\le $ 100KB), they only account for around 5\% of total bytes transferred \cite{Alizadeh-2010-Data,Alizadeh-2013-pFabric}. Therefore the overhead of replicating mice flows is rather small, and the negative impact rather limited. In addition, RepFlow is transport agnostic, and works with legacy TCP as well as newer protocols such as DCTCP \cite{Alizadeh-2010-Data}. It can be implemented as a general library or middleware for any application to use. Through queueing analyses and trace-driven simulations, Xu et al. demonstrate that RepFlow provides 50\%--70\% speedup in both mean and 99-th percentile FCT for all loads over TCP. 




\section{Challenges \& Opportunities}
\label{sec:discussion}

Research in low latency datacenter networking is still in an early stage with many immediate opportunities. Yet, some fundamental challenges remain open. We believe that first, from a theory perspective, a complete model of end-to-end latency taking into account the characteristics of datacenter networks is fundamentally interesting. It will, to a great deal, help us systematically understand the design space and the tradeoffs to construct better protocols. Second, from a practical perspective, another fundamental challenge is how to provide a simple network abstraction for the applications, with meaningful and explicit latency performance guarantees. If we can orchestrate our datacenter network so that it guarantees the mean and tail FCT for a flow of given size, even just in a statistical sense, the application design can be greatly simplified without the hairy details of coping with the unpredictable networks. 

The deployment of new technologies in datacenter networks may also inspire brand-new research directions of low latency. First, 3D-beamforming \cite{ZZZL12, Hamedazimi-2014-FireFly} have been taken as a cure for the bursty traffic. With these wireless links which provide additional flexibility, reliability and scalability, long latency caused by hotspots may be mitigated given that a practical scheduler can be designed to adapt to the changing topology. Second, arming congestion control mechanisms with learnability \cite{Sivaraman-2014-Learnability} is likely to be a promising solution to lower latency amid dynamic and complicated network conditions.

\section{Conclusion}
\label{sec:conclusion}

We have surveyed, to our best effort, a small sample of recent proposals on reducing FCT for mice flows in datacenter networks. Four techniques, namely reducing queue length, accelerating retransmissions, prioritizing mice flows, and utilizing multipath, emerged as effective means to alleviate long queueing delay and packet losses. Looking forward, challenges and opportunities co-exist beyond the examples we outlined. We thus encourage a broader and deeper investigation into this problem from not only the networking community, but also other communities as well as the industry.

\bibliographystyle{abbrv}
\bibliography{main}

\end{document}